\documentstyle[11pt,newpasp,twoside,epsf]{article}

\newcommand{\ds}{\displaystyle}

\begin{document}
\title{Mode identification from line-profile variations}
\author{C. Aerts\altaffilmark{1} \& L. Eyer}
\affil{Instituut voor Sterrenkunde, Katholieke Universiteit Leuven, Belgium\\
conny@ster.kuleuven.ac.be; laurent@ster.kuleuven.ac.be}
\altaffiltext{1}{Postdoctoral Fellow, Fund for Scientific Research, Flanders}

\begin{abstract}
We review the current status of different mode-identification techniques based
on observed line-profile variations. Three basic methods are currently
available to identify the non-radial pulsation modes. These three methods are
described, together with their different variants. We further present
applications to real data, focusing especially on $\delta\,$Scuti stars.
After having discussed all the methods and their applications, we present an
inventory in which we compare the properties of the different methods. Finally,
we end with future prospects, both on the theoretical and observational side.
\end{abstract}


\section{Introduction}

Intrinsic variable stars are an important diagnostic to test stellar models,
which provide in turn valuable clues to our understanding of stellar and
galactic evolution. Variable stars pulsate in so-called {\it non-radial
pulsation modes}.  Since the early seventies, ample observational evidence of
the presence of such non-radial pulsations has become available.  From then on,
detailed observational and theoretical studies of non-radial pulsations have
been conducted. It has become clear that by understanding the pulsations in
full detail, one can probe the internal structure of the stars and hence
confront the results with stellar evolution theories, i.e., one can apply {\it
asteroseismology}. 

The non-radial pulsations lead to periodic variations of physical quantities,
such as surface brightness, radial velocity, temperature, etc.  By comparing
the observed variations with those predicted by theory, it is in principle
possible to determine the most important parameters that characterise the
pulsation.  One specific aspect of studying pulsations therefore is to know
what kind of modes are active in pulsating variables, i.e., the aspect of {\it
mode identification}.  Specifically, mode-identification techniques try to
assign values to the spherical wavenumbers $(\ell,m)$: the degree and azimuthal
number of the spherical harmonic $Y_{\ell}^m$ that describes the non-radial
pulsation.  The identification of non-radial pulsation modes from observational
data of variable stars is important, since it is a first step towards
asteroseismology.  Indeed, the amount of astrophysical information that can be
derived from the study of non-radial pulsations depends directly on the number
of modes that can be successfully identified.

Inspired by this potential of  identifying the non-radial pulsation modes, the
study of {\it mode-identification techniques} has become an extended topic by
itself in variable star research. We have been involved in the development and
the application of such mode-identification methods.  In this review, we focus
on the different identification techniques that are currently used to identify
non-radial pulsations from line-profile variations.

The introduction of high-resolution spectrographs with sensitive detectors has
had an enormous impact on the field of mode identification.  Spectroscopic data
usually offer a very detailed picture of the pulsation velocity field. On the
other hand, they require large telescopes and sophisticated instrumentation.
Before accurate detectors were available, identifications had to be obtained
from photometric observations.  These kind of data are suitable to study
long-period pulsations because they can be obtained with small telescopes,
which are available on longer time scales. For a review on photometric studies
of $\delta\,$Scuti stars we refer to Poretti and to Garrido (these
proceedings).

The plan of our paper is as follows. We first briefly give a (non-mathematical)
introduction into the domain of non-radial pulsations. Next, we explain how a
theoretical line profile can be calculated for a non-radially
pulsating star.  The following section is devoted to the description of the
different identification techniques: line-profile fitting, Doppler Imaging, and
the moment method. We briefly review the main characteristics of each method in
Section\,5 and we finally end with some future prospects in this
field of research. In particular, we point out the importance of the
identification of pulsation modes in $\gamma\,$Dor stars.

\section{Non-radial pulsations}
With a radial pulsation the physical parameters throughout a star vary
periodically along the radial direction and spherical symmetry is preserved
during a pulsation cycle. The differential equation describing the radial
displacement is of the Sturm-Liouville type and thus allows eigensolutions that
correspond to an infinitely countable amount of eigenfrequencies.  The smallest
frequency corresponds to the fundamental radial pulsation mode. The period of
this mode is inversely proportional to the square root of the mean density of
the star. Radial pulsations are characterised by the radial wavenumber $n$: the
number of nodes of the eigenfunction between the center and the surface of the
star.
 
If transverse motions occur in addition to radial motions, one uses the term
non-radial pulsations.  The pulsation modes are then not only characterised by
a radial wavenumber $n$, but also by non-radial wavenumbers $\ell$ and $m$. The
latter numbers 
correspond to the degree and the azimuthal number of the spherical
harmonic $Y_{\ell}^m$ that represents the dependence of the mode on the angular
variables $\theta$ and $\varphi$ for a star with a spherically symmetric
equilibrium configuration (see Equation\,(\ref{nrp}) below).  The degree $\ell$
represents the number of nodal lines, while the azimuthal number $m$ denotes
the number of such lines that pass through the rotation axis of the star.
For stars that
are not spherically symmetric, the expansion of the eigenfunctions in terms of
spherical harmonics is no longer obvious.

A distinction is made between $p$-modes, $g$-modes, and $r$-modes.  In $p$-modes, the
restoring force is the pressure force; radial modes can be viewed as a special
case of non-radial $p$-modes. In $g$-modes, the restoring force is the buoyancy
force; such modes have periods that are longer than the period of the radial
fundamental mode. Finally, $r$-modes or toroidal modes are characterised by
purely transverse motions; such modes only attain finite periods in rotating
stars.

In the case of spheroidal modes in the approximation of a non-rotating star,
the pulsation velocity expressed in a system of spherical coordinates
$(r,\theta,\varphi)$ centered at the centre of the star and with polar axis
along the symmetry axis of pulsation, is given by
\begin{equation}
\label{nrp}
\ds{\vec{v}_{\rm puls}=
\left(v_r,v_{\theta},v_{\varphi}\right)=N_{\ell}^mv_{\rm p}
\left(1,K\frac{\partial}{\partial\theta},\frac{K}{\sin\theta}
\frac{\partial}{\partial\varphi}\right)Y_{\ell}^m(\theta,\varphi)
\exp{\left({\rm i}\omega t\right)}
}
\end{equation}
(e.g., see Smeyers 1984, Unno et al.\ 1989).
In this expression, $N_{\ell}^m$ is the normalisation factor for the
$Y_{\ell}^m(\theta,\varphi)$ over the visible hemisphere of the star, $v_{\rm
p}$ is the pulsation amplitude, $\omega$ is the pulsation frequency, 
and $K$ is the ratio of the horizontal to the vertical velocity amplitude. The
latter can be found from the boundary conditions: $K=GM/(\omega^2R^3)$. 
The sign of the azimuthal number $m$ describes how the mode progresses with
respect to the rotation of the star. We here adopt the convention that positive
$m$-values represent waves that travel opposite to the rotation (retrograde
modes), while negative $m$-values are associated with modes that travel in the
direction of the rotation (prograde modes). Modes with $m=0$ are axisymmetric
modes, while those with $\ell=|m|$ are called sectoral modes. In all other
cases ($\ell\neq |m|$ and $m\neq 0$) one speaks of tesseral modes.

A caveat for many analyses on non-radial pulsations is that the theoretical
framework that is used only applies in the slow-rotation approximation, i.e.,
in the case where the effects of the Coriolis force and of the centrifugal
forces can be neglected in deriving an expression for the components of the
pulsation velocity.  We emphasize that it is not allowed to describe an
oscillation mode for a rotating star in terms of a single spherical harmonic,
and so to ascribe a single set of wavenumbers ($\ell,m$) to a mode.  The
Coriolis force, for instance, introduces a transverse velocity field that is of
the same order of magnitude as the pulsation velocity for a non-rotating star
if the ratio $\Omega/\omega$ of the rotation frequency to the pulsation
frequency approaches unity. In particular, we have studied the effect of the
Coriolis force on line-profile variations in the 
case of $p$-modes (Aerts \& Waelkens
1993) and we have found that the line profiles can be largely influenced for
some stellar parameters.  A comparable study for $g$-modes was presented
by Lee \& Saio (1990).  For stars having $\Omega/\omega\approx 1$ or larger,
the centrifugal forces also become important. Including the latter enormously
increases the complexity of the mathematical treatment of the problem, because
deviations from spherical symmetry have to be taken into account. It is clear
that $\Omega/\omega$-values, which are too large to be neglected if the aim is
to obtain an accurate description of the pulsation, are met in several stars
discussed in the literature. A re-evaluation of observed line profiles in rapid
rotators is therefore necessary in some cases.

\section{Line-profile variations}

The velocity field caused by the non-radial pulsation(s) leads, through Doppler
displacement, to periodic variations in the profiles of spectral lines.
Theoretical line-profile variations can be calculated in the following way.
Consider a system of spherical coordinates $(r,\theta,\varphi)$ with the polar
axis coinciding with the direction to the observer.  The velocity field due to
the rotation and the pulsation leads to a Doppler shift at a point
$(R,\theta,\varphi)$ on the visible surface of the star.  The local
contribution of a point $(R,\theta,\varphi)$ to the line profile is
proportional to the projected intensity of that
point. We approximate this projected intensity as follows.
We divide the stellar surface into a number of 
infinitesimal surface elements, which, for computational purposes, have 
finite dimensions. Next, we assume that
the intensity  $I_{\lambda}(\theta,\varphi)$ 
is the same for all points of the considered surface element.
The projected intensity of the surface element
surrounding the point $(R,\theta,\varphi)$ then
is the product of the intensity  $I_{\lambda}(\theta,\varphi)$ 
and the projection on the line
of sight of the surface element around the considered point:
\begin{equation}
I_{\lambda}(\theta,\varphi)R^2\sin\theta\cos\theta\ d\theta\ d\varphi.
\end{equation}

Because of variations of the intensity over the stellar surface, and of the
temperature dependence of an absorption line, the contributions of the
different points on the visible stellar surface to the line profile have a
different amplitude.  
In first instance, however, one assumes that 
the perturbations of the intensity and of the
surface affect the line profile only slightly. These effects are therefore
neglected and it is assumed that
\mbox{$\delta I_{\lambda}(\theta,\varphi)=0$} during the pulsation. 
The time dependence of the intensity is important when the spectral 
line
is sensitive to the temperature and when the temperature differs for different
points on the stellar surface.  This time dependence is also neglected in most
calculations.

The most important effect that then changes the projected intensity over the
visible surface is the limb darkening.  Usually, the intensity is assumed to be
isotropic in the $\varphi$ coordinate and 
the $\theta$-dependence of the
intensity is described by a limb-darkening law of the form
\begin{equation}
\label{randverduistering}
h_{\lambda}(\theta)=1-u_{\lambda}+u_{\lambda}\cos\theta,
\end{equation}
where $u_{\lambda}\in [0,1]$ is called the limb-darkening coefficient; it
depends on the considered wavelength range. Wade \& Rucinski (1985) have
tabulated limb-darkening coefficients in terms of temperature, gravity
and wavelength.
The projected intensity of a surface element centered around the point
$P(R,\theta,\varphi)$ with size $R^2\sin\theta\ d\theta\ d\varphi$ then is
\begin{equation}
\label{energie}
I_0h_{\lambda}(\theta)R^2\sin\theta\cos\theta\ d\theta\ d\varphi,
\end{equation}
where $I_0$ is the intensity at $\theta=0$.

In order to take into account intrinsic broadening effects, the local line
profile is convolved with an intrinsic profile, which we take to be Gaussian
with variance ${\sigma}^2$, where $\sigma^2$ depends on the spectral line
considered. Generalisations to an intrinsic Voigt profile or a profile derived
from a stellar atmosphere model are easily performed.

Let us represent by $p(\lambda)$ the line profile and by ${\lambda}_{ij}$ the
Doppler-corrected wavelength for a point on the star with coordinates
$({\theta}_i,{\varphi}_j)$, i.e.,
\begin{equation}
\ds{\frac{{\lambda}_{ij}-{\lambda}_0}{{\lambda}_0}=
\frac{\lambda ({\theta}_i,{\varphi}_j)-{\lambda}_0}{{\lambda}_0}=
\frac{\triangle \lambda({\theta}_i,{\varphi}_j)}{{\lambda}_0}=
\frac{v({\theta}_i,{\varphi}_j)}{c}}
\end{equation}
with $v({\theta}_i,{\varphi}_j)$ the component of the 
sum of the pulsation and rotation velocity
of the considered point in the line of sight. An explicit expression for
$v({\theta}_i,{\varphi}_j)$ can be found in e.g., Aerts et al.\ (1992).  The
line profile can then be approximated as
\begin{equation}
\label{conv}
\ds{p(\lambda)=\sum_{i,j}
\frac{I_0h_{\lambda}({\theta}_i)}
{\sqrt{2\pi}\sigma}
\exp{
\left(
{-
\frac{(\lambda_{ij}-\lambda)^2}{2\sigma^2}
}
\right)
} R^2\sin\theta_i\cos\theta_i\triangle\theta_i\triangle\varphi_j},
\end{equation}
where the sum is taken over the visible stellar surface
($\theta\in[0^{\circ},90^{\circ}], \varphi\in[0^{\circ},360^{\circ}[$).
\begin{figure}
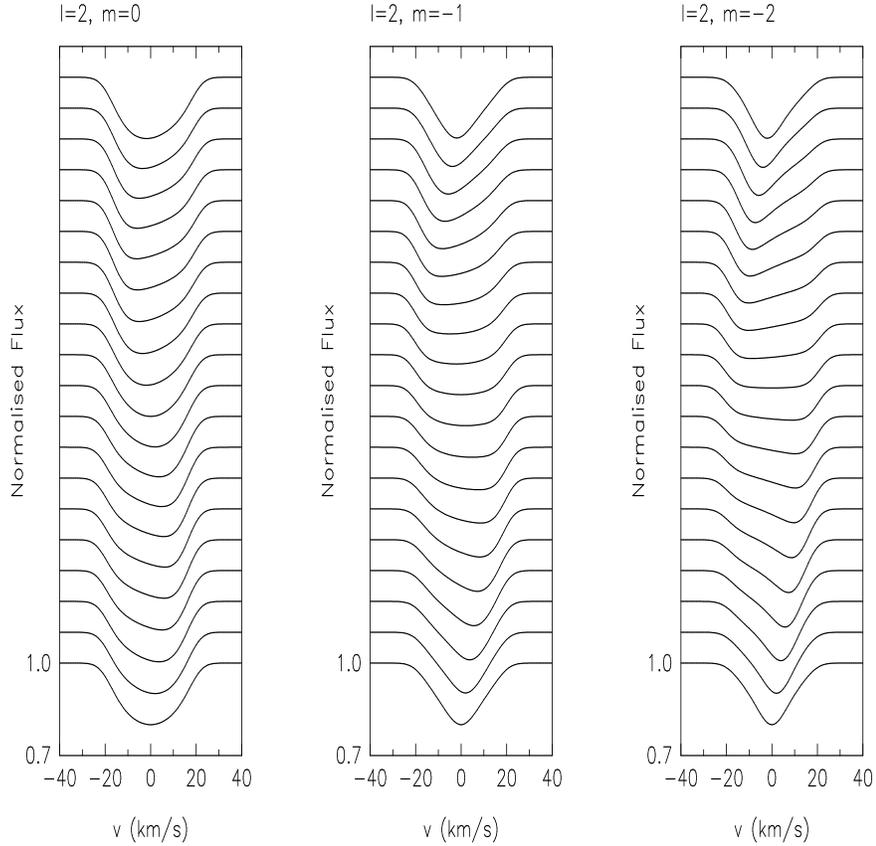

\mbox{\epsfxsize=.3\textwidth\epsfysize=0.9\textwidth\epsfbox[0 0 250 500]
{aerts1.ps}}
\mbox{\epsfxsize=.3\textwidth\epsfysize=0.9\textwidth\epsfbox[0 0 250 500]
{aerts2.ps}}
\mbox{\epsfxsize=.3\textwidth\epsfysize=0.9\textwidth\epsfbox[0 0 250 500]
{aerts3.ps}}
\caption{Theoretically determined line-profile variations calculated by means of
the basic formalism given in Section\,3. We used an $\ell=2$ mode and 
respectively
$m=0$ (left panel), $m=-1$ (middle panel), and $m=-2$ (right panel). The other
velocity parameters are: $v_{\rm p}=5\,$km/s, $v\sin\,i=30\,$km/s, 
$\sigma=4\,$km/s, and $i=55^{\circ}$. The phase of each profile increases from 
0.0 (lowest profile) to 1.0 (highest profile) in steps of 0.05. }
\end{figure}

We show in Figures\,1 and 2  sets of theoretically calculated profiles for
$\ell=2$ and $\ell=6$ modes. The profiles in Figure\,1 are for prograde modes, 
those in
Figure\,2 for retrograde modes. The other velocity parameters are 
$v_{\rm p}=5\,$km/s, $v\sin\,i=30\,$km/s, $\sigma=4\,$km/s, and $i=55^{\circ}$.
Other studies in which theoretical profiles are given are published by Kambe \&
Osaki (1988) and by Schrijvers et al.\ (1997).

\begin{figure}
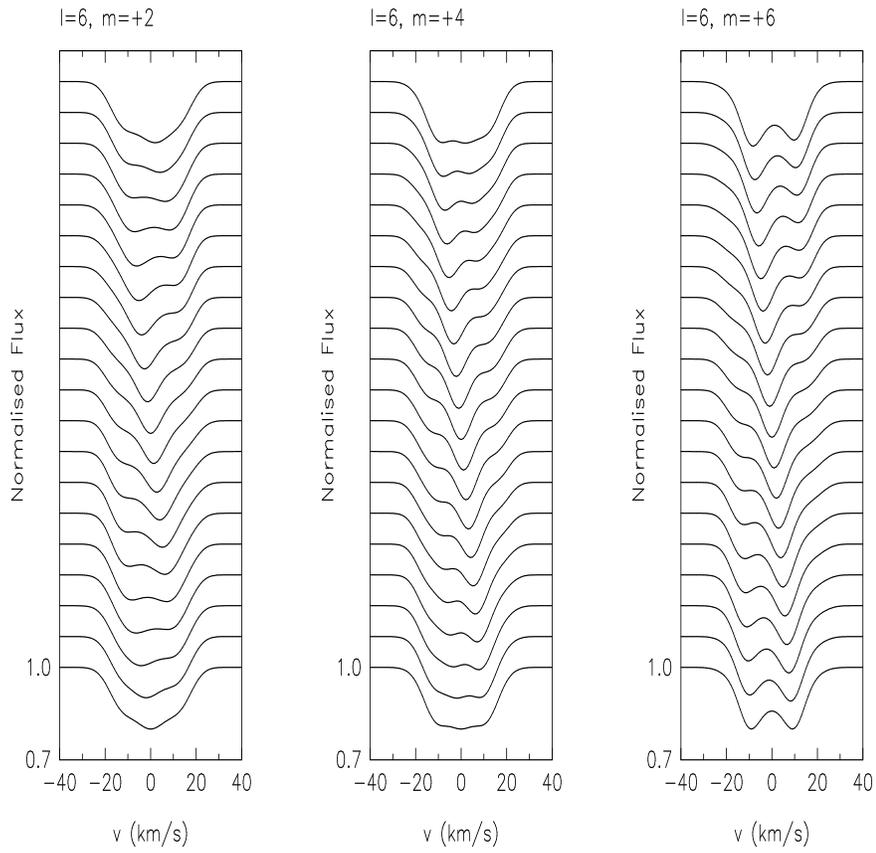

\mbox{\epsfxsize=.3\textwidth\epsfysize=0.9\textwidth\epsfbox[0 0 250 500]
{aerts4.ps}}
\mbox{\epsfxsize=.3\textwidth\epsfysize=0.9\textwidth\epsfbox[0 0 250 500]
{aerts5.ps}}
\mbox{\epsfxsize=.3\textwidth\epsfysize=0.9\textwidth\epsfbox[0 0 250 500]
{aerts6.ps}}
\caption{The same as in Figure\,1, but for 
$\ell=6$ with $m=+2$ (left panel), 
$m=+4$ (middle panel), and $m=+6$ (right panel).}
\end{figure}

Ideally, the calculation for the line profile described above should be
generalised in order to take into account the following additional
time-dependent effects: a variable surface normal, a variable intensity through
non-adiabatic temperature variations, a variable intrinsic profile, Coriolis
and centrifugal correction terms to the pulsation velocity expression. The most
up-to-date code that takes into account some of these effects is written by
Townsend (1997).  He used Lee et al.\ 's (1992) formalism to take into account
rotation effects.  This formalism incorporates the Coriolis force  for all
values of $\Omega/\omega$, but neglects the centrifugal forces, which are $\sim
\Omega^2$. A variable surface normal is taken into account, but the intensity
variations are still assumed to be adiabatic according to the approximation
presented by Buta \& Smith (1979). This user-friendly code published by
Townsend (1997) is available upon request from the author.

\section{Identification techniques}

In this section, we describe the different methods that are used to identify
modes. It is clear that the velocity expression based on the non-radial
pulsation model presented above contains many free parameters, even in the
simple formulation in which rotational and non-adiabatic effects are neglected.
Especially the infinity of candidate modes is a problem when constructing
identification techniques and it often keeps the predictive power of the
methods low. This is in particular the case for the methods that are based on a
trial-and-error principle. We point out that quantitative methods are better to
obtain a reliable mode identification.  This need for quantitative methods has
become apparent since more and more detailed spectroscopic analyses have
revealed that multimode pulsations are often present.  Below, we treat three
methods, more or less in the order of their appearance in the literature.

In describing the methods, we assume that the pulsation frequencies have been
determined from the observables of the variable stars. For a description of the
different methods used to derive the modal frequencies, we refer to the review
of Mantegazza: Mode detection from line-profile variations (these proceedings).
We pay special attention to describe applications to $\delta\,$Scuti stars in
this paper. For a review on identification methods applied to OB-type variables
we refer to Aerts (1994).

\subsection{Objective line-profile fitting}

Since Osaki (1971) computed theoretical line profiles for various non-radial
pulsations, the identification of modes from spectroscopic observations has
become possible.  The identification of non-radial pulsation modes from
line-profile variations was first achieved by line-profile fitting on a
trial-and-error basis.  The idea is to compare the observed line-profile
variations with those predicted by theoretical calculations.  This technique
was the first one in use to identify modes from spectroscopic observations.
\begin{figure}
\mbox{\epsfxsize=1.07\textwidth\epsfysize=1.07\textwidth\epsfbox[75 200 570 700]
{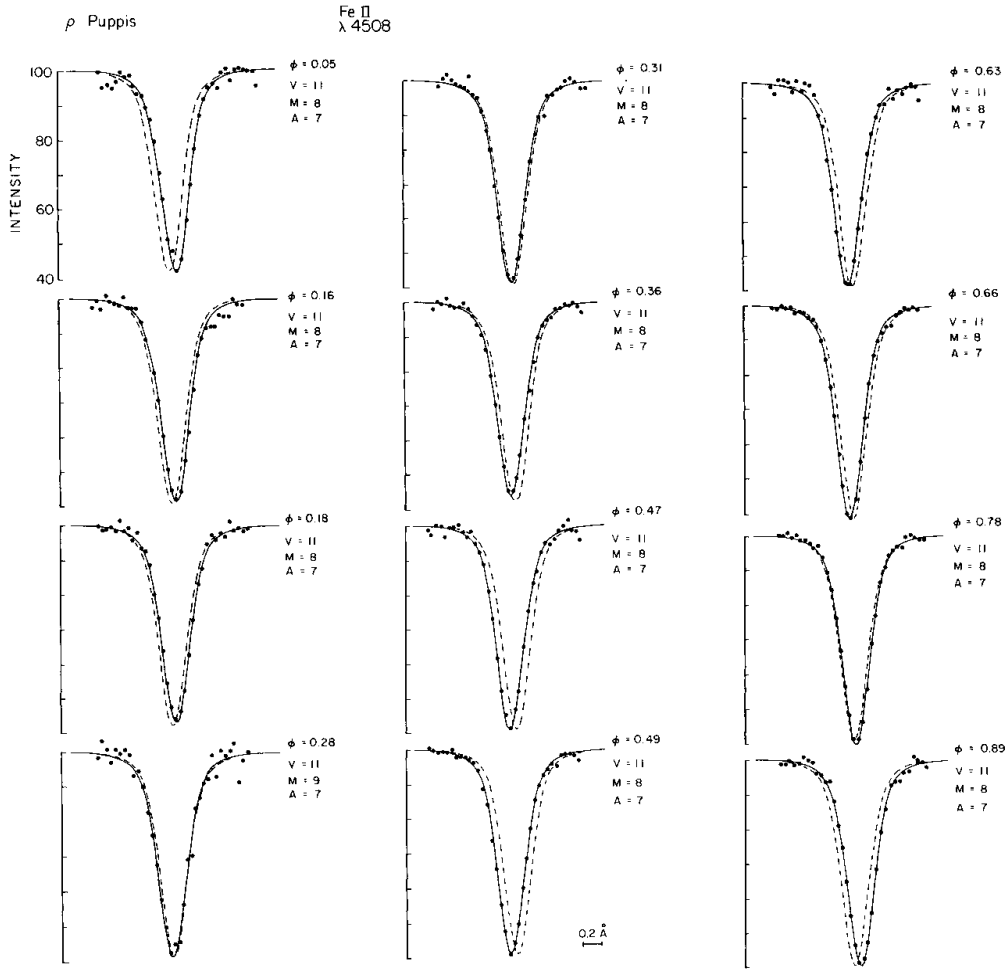}}
\caption{Observed (dots) and theoretical line-profile variations of the
$\delta\,$Scuti star $\rho\,$Puppis. The full line is a model for a radial
pulsation while the dashed line is a model for which only rotational broadening
appears. V stands for the projected rotation velocity, M for the intrinsic
broadening, and A for the amplitude of the pulsation; these three velocities
are indicated next to each panel and are expressed in km/s.  Figure taken with
permission from Campos \& Smith (1980b).}
\end{figure}
\begin{figure}
\mbox{\epsfxsize=1.\textwidth\epsfysize=1.\textwidth\epsfbox[150 20 570 520]
{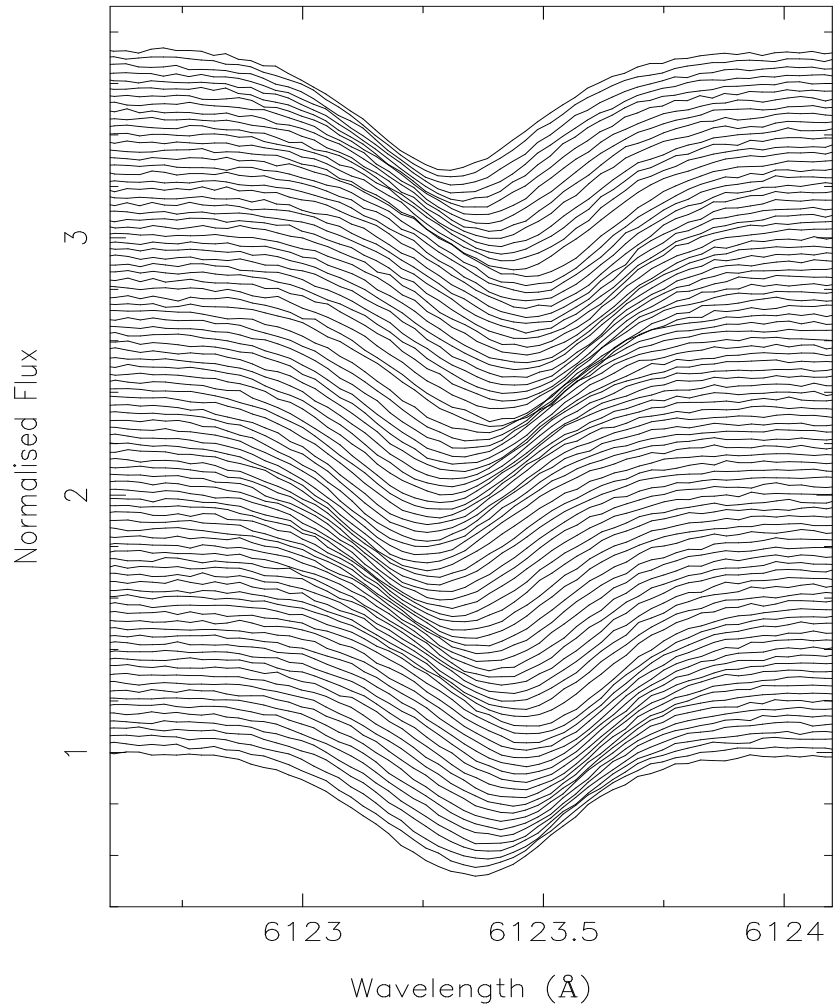}}
\caption{Line-profile variations of the $\delta\,$Scuti
star $\rho\,$Puppis observed with the CAT/CES of ESO by Mathias et al.\ (1997).
 The quality of these data are much better than those
presented in Figure\,3. They have revealed the presence of
additional small-amplitude modes, besides the main radial mode.}
\end{figure}

Pioneering work in the field of line-profile fitting was done by M. Smith and
his collaborators. They obtained line profiles for various types of pulsating
stars.  We show in Figure\,3 their observed line profiles,
represented as dots, of the $\delta\,$Scuti star $\rho\,$Puppis (Campos \&
Smith 1980b).  The theoretical profiles that are presented by the full line
are constructed with a radial mode. The dashed curves represent rotationally
broadened lines, i.e., lines for a non-pulsating star. These dashed lines show
that the line profiles of $\rho\,$Puppis are indeed variable in time because of
the pulsation of the star.  The representation by the full line is rather
faithful and the authors concluded that the star pulsates radially.  Later on,
it was found by means of the moment method applied to more recent spectra (see
Figure\,4 and Section\,4.3) that the main mode is indeed
radial, but that at least one, and probably two, additional small-amplitude
modes are present in $\rho\,$Puppis (Mathias et al.\ 1997), a conclusion
that would have been very hard to obtain by means of the fitting technique.

In the seventies and early eighties, the fitting technique by trial-and-error
was very popular, simply because it was the only one available.  Besides
applications in the case of $\delta\,$Scuti stars (Campos \& Smith 1980b, Smith
1982), the technique was also applied to line-profile variations of
$\beta\,$Cephei stars (Smith 1977, 1983; Campos \& Smith 1980a), Be stars
(e.g., Vogt \& Penrod 1983, Baade 1984), and the so-called 53\,Per stars (Smith
1977, Smith et al.\ 1984). Smith (1982) describes mode-typing for 9
$\delta\,$Scuti stars by means of line-profile fitting. Most of the results
that he obtained are still valid today as far as the main modes of the stars
are concerned.

As already mentioned, the main disadvantage of the trial-and-error line-profile
fitting is the large number of free parameters that appear in the velocity
expression due to the non-radial pulsation. In principle, the complete free
parameter space has to be scanned before a decision on the best mode can be
obtained.  This was not yet possible some 15 years ago, because it was
computationally too demanding. For this reason, only a limited number of
combinations were tried out, with the result that the identification technique was not very
objective.  Moreover, the non-radial pulsation model can be quite successful in
reproducing the line profiles observed on a short time scale 
for different sets of input parameters, i.e., the
fitting technique does not necessarily lead to a unique solution.  We also
point out (see Aerts et al.\ 1992, Aerts \& Waelkens 1993) that the apparent
quality of some fits is suspect in the sense that in early modeling, one
neglected temperature variations and rotational effects, which obviously must
affect the profiles in some cases (Lee et al.\ 1992).

Other problems that appeared in early applications of the line-profile-fitting
technique are caused by the often very limited time base of the data, because
of which it was sometimes necessary to assume that modes temporarily disappear
in order to re-obtain good fits for new data that span a longer time scale.
Also, the values found for the intrinsic profile sometimes had to be varied
from one night to another in order to obtain reliable fits.  In the case of
rapidly rotating stars, one usually assumed equator-on geometries and
high-degree sectoral modes because these are the ones that best reproduce the
observed moving bump phenomenon.  Finally, it was mentioned, but most often not
taken into account for applications to rapid rotators, that one used an
expression for the pulsation velocity that is related to one spherical
harmonic. This is, however, only valid in the case of a non-rotating star. All
the abovementioned assumptions were introduced in an {\it ad hoc} fashion and
cast doubt on the reliability of the model.

Nowadays, it is possible to identify the pulsation mode by performing
line-profile fitting in an objective way. This can be achieved by calculating a
kind of overall standard deviation per wavelength pixel between theoretically
determined and observed profiles for a large grid of possible wavenumbers and
velocity parameters. In order to do so, one needs a large homogeneous data set
of high-resolution profiles that are well-spread over the period that appears
in the line-profile variability.  The theoretical limitations of the model have
also mostly been overcome by now, as explained in Section\,3.

A plus point of objective line-profile fitting is that both the wavenumbers
($\ell,m)$ and all the other velocity parameters are derived. In this way, the
complete motion due to pulsation can be reconstructed once the best fit is
selected. 

THE major drawback of objective line-profile fitting is that it is still
limited to a monoperiodic pulsation. Indeed, it is in practice impossible to
consider combinations of all kinds of different modes to fit the data without
any restriction on the parameters. It is nevertheless useful to use fitting for
multiperiodic stars, once estimates of the spherical wavenumbers and the
velocity parameters are at hand from other methods such as those presented in the
following two sections.

\subsection{Doppler Imaging}

In recent spectroscopic studies,  a lot of attention has been paid to the
line-profile variations of rapidly rotating OB stars. This has been in
particular the case since it was recognised that the line profiles of rapid
rotators allow a Doppler Imaging of the stellar surface (Vogt et al.\
1987), so that a mapping of the velocity over this surface is possible (Baade
1987).

\begin{figure}
\mbox{\epsfxsize=0.9\textwidth\epsfysize=0.9\textwidth\epsfbox[30 170 550 730]
{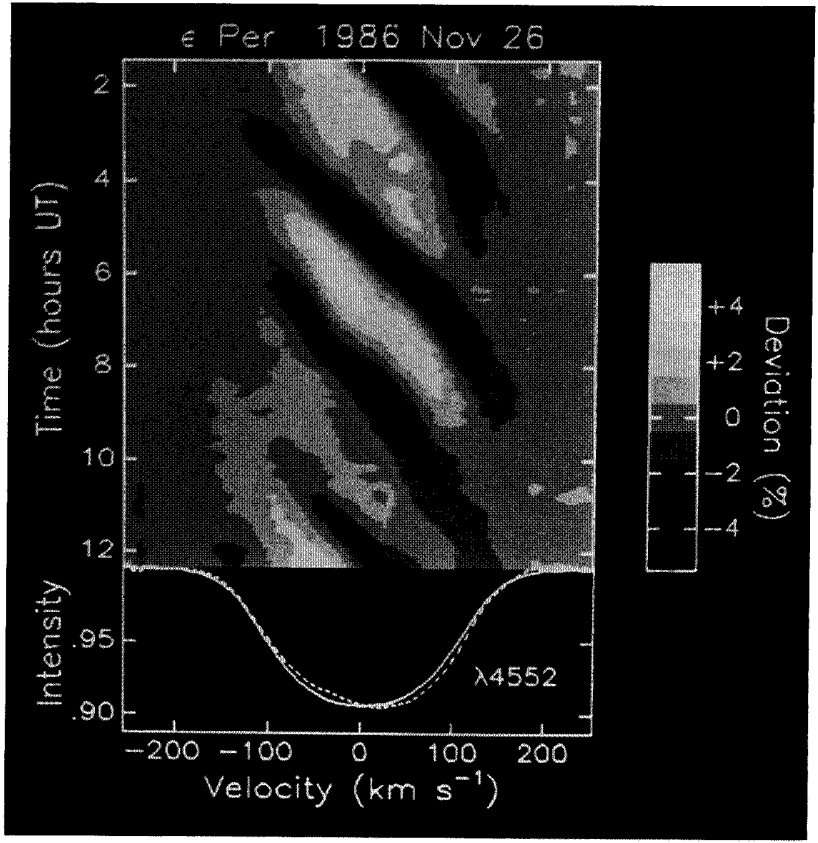}}
\caption{Grey-scale representation of the observed line-profile variations of
$\varepsilon\,$Per. The spectra are residuals with respect to the global
average line profile, shown as full line in the lower panel. The dashed line is
the nightly average profile. Darker shades indicate a depression relative to
the mean line while bright regions correspond to places where the profile is
shallower than the mean. Figure taken from Gies \& Kullavanijaya (1988) with
permission.}
\end{figure}

Gies \& Kullavanijaya (1988) first presented an objective criterion based on
Doppler Imaging to determine the periods and pulsation parameters of the modes
in the rapidly rotating line-profile variable B\,0.7\,III star
$\varepsilon\,$Per.  In Figure\,5 we show a grey-scale representation of the
residual line-profile variations (with respect to the global symmetric line
profile) of $\varepsilon\,$Per obtained on 26 November 1996. Black denotes
local deficiencies of the flux and white local increments. Gies \& Kullavanijaya
noted that emission patterns that move through the line profile during the
pulsation cycle are easily detected and visualised in such a representation.
This way of presenting data has since then become very popular.  Fourier
analysis of the line-profile variations at each wavelength point yields the
periods of the variations by frequency peaks in the resulting periodogram.

Subsequently, the azimuthal number $m$ is obtained by considering the number of
phase changes $\triangle$(Phase) 
at each signal frequency versus the line position. These observed
phase changes are shown in Figure\,6 in the case of the four
frequencies detected by Gies \& Kullavanijaya in their line-profile variations
of $\varepsilon\,$Per. The basic idea behind the estimation of $m$ is the
following. Let us {\bf assume} that sectoral modes are excited, that we are
dealing with an equator-on view and that the bump motion is caused by the large
rotation of the star. Since each of the three components of $\vec{v}_{\rm puls}$
is proportional to $\exp{\left({\rm i}(\omega t + m\varphi)\right)}$,
the phase decreases by $m/2$ cycles between the blue and the red wing of the
profile. In this way, an upper limit of $m$ is given by 2$\triangle$(Phase).
On the other hand,
\begin{equation}
\ds{\frac{d{\rm Phase}/d\varphi}{dV_{\rm rot}/d\varphi}=
\frac{m/2\pi}{V_{\rm eq}\sin\,i\sin\theta\cos\varphi}},
\end{equation}
where $V_{\rm rot}$ is the component of the rotation velocity in the
line-of-sight and $V_{\rm eq}$ stands for the equatorial rotation speed.
In this way,
\begin{equation}
2\pi (V_{\rm eq}\sin\,i)
\ds{\frac{d({\rm Phase})}{dV_{\rm rot}}}
\end{equation}
is a lower limit for $m$.  By calculating both limits, one obtains an estimate
of the azimuthal number of the mode.

\begin{figure}
\begin{center}
\mbox{\epsfxsize=0.78\textwidth\epsfysize=0.78\textwidth\epsfbox[300 135 565 577] 
{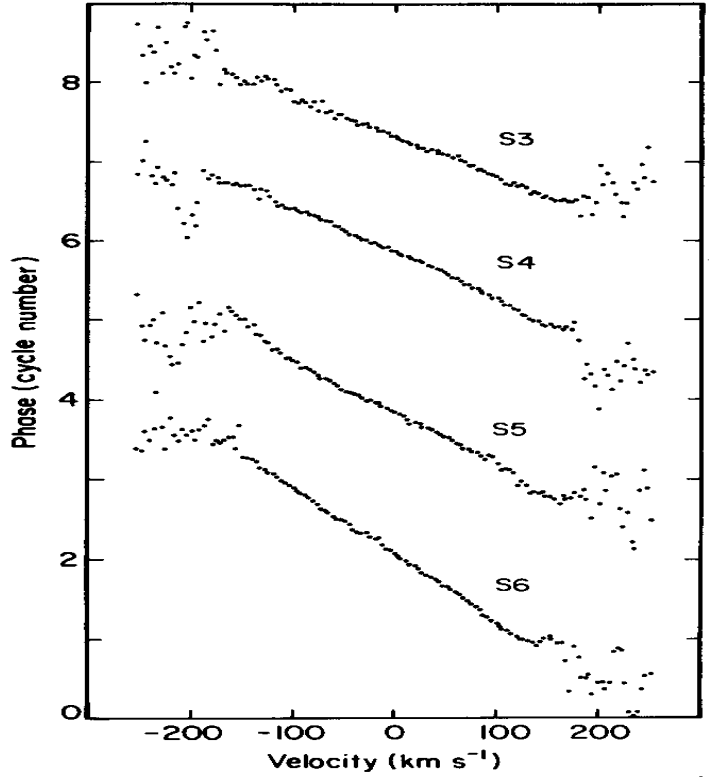}} 
\end{center}
\caption{
The phase of the power spectrum of the line-profile variations of
$\varepsilon\,$Per as a function across the line profile. Each plot corresponds
to the phases at the peak frequencies S3, S4, S5, S6 that were found in the
profile variations. An estimate of the azimuthal number $m$ is derived from
these phase changes (see text for an explanation).  Figure taken from Gies \&
Kullavanijaya (1988) with permission.}
\end{figure}
\begin{table}
\caption{The limits for the azimuthal number $m$ for the star
$\varepsilon\,$Per as derived by Gies \&
Kullavanijaya (1988)}
\begin{center}
\tabcolsep=8pt
\begin{tabular}{cccc}
\tableline
{Signal}&{Lower Limit}&{Upper limit}&{Adopted}\\
\tableline
S3 & -3.93 & -2.98 & -3 \\
S4 & -4.42 & -3.60 & -4 \\
S5 & -5.15 & -4.46 & -5 \\
S6 & -6.90 & -5.20 & -6
\\
\hline
\end{tabular}
\end{center}
\end{table}
In Table\,1 we list the limiting values for the azimuthal number $m$ obtained
by Gies \& Kullavanijaya (1988) from the observed phase changes shown in
Figure\,6. Additional frequencies and interpretations of the line-profile
variability of $\varepsilon\,$Per 
are available by now (see e.g., Gies et al.\ 1999), but we do not
want to describe the details here since the $\varepsilon\,$Per case was only
given as an example of the method.

A major disadvantage of the Doppler Imaging method as proposed by Gies \&
Kullavanijaya (1988) is that equator-on-viewed sectoral modes are assumed
without a real physical argument.  For this and also other reasons, a number of
generalisations of the method have been proposed in the literature. Merryfield
\& Kennelly (1993) propose to use the Doppler Imaging principle to obtain the
wavenumbers by considering a two-dimensional Fourier transform, which leads to
power diagrams as a function of the frequency and as a function of 
what they call the
``apparent'' azimuthal number \it{\^m}\rm. They propose that
\it{\^m}\rm\ may be 
an estimate of the degree rather than the azimuthal number in
the case of tesseral modes. This finding was unambiguously proven 
for the first time for all
considered modes by Telting \& Schrijvers (1997), who performed an extensive
simulation study to evaluate the Doppler Imaging method as an identification
method. They also found that the phase variation for the first harmonic of the
frequency contains information on the azimuthal number $m$. So far, Schrijvers 
\& Telting applied their method to (new) $\beta\,$Cep stars 
(for an overview of their applications, see Schrijvers 1999).

Kennelly \& collaborators (Kennelly et al.\ 1992a: $\tau\,$Peg; 1992b:
$\gamma\,$Boo; 1996: $\theta^2\,$Tau; 1998a: $\varepsilon\,$Cep) 
are the only ones
so far who have actually used the Doppler Imaging principle to obtain the
wavenumbers for a group of rapidly rotating $\delta\,$Scuti stars. Hereto, they
gathered numerous high-quality spectra and they considered a two-dimensional
Fourier transform.  Their latest technique consists of the following two steps:
\begin{itemize}
\item
perform a DDLPV: Doppler Deconvolution of line-profile variations. First, one
derives the intrinsic profile $\psi(v)$ from  the deconvolution
${\overline\phi} (v) = R(v) \ast
\psi(v)$, where  ${\overline\phi} (v)$ is the time-averaged line profile and
$R(v)$ the rotationally broadened profile. As first guess for $\psi$, the
synthetic spectrum from a model atmosphere is taken. Subsequently, the observed
time-dependent pulsationally broadened components $\phi(v,t)$ of the spectra
are modeled from the deconvolution $\phi(v,t) = B(v,t) \ast \psi(v)$. As
initial guess for $\phi$ they take the rotational broadening $R$.

In Figure\,7 we show the result of the DDLPV technique applied by
Kennelly et al.\ (1998a) to their line-profile variations of the $\delta\,$Scuti
star $\varepsilon\,$Cep.
\item
perform FDI: Fourier Doppler Imaging, by remapping the time-variable component
of $B(v,t)$  from velocity to Doppler space $B(\phi,t)$ by means of
$\phi_i=\sin^{-1} (v_i/v\sin\,i)$. Next, the two-dimensional Fourier transform
of $B(\phi,t)$, and the corresponding amplitude spectrum, is computed.
\end{itemize}
We refer to Kennelly et al.\ (1998b) for more details, but we point out here
that $\overline{\phi}(v)$ also contains a broadening component due to the
pulsation which is not taken into account by the authors.  

\begin{figure}[t]
\mbox{\epsfxsize=0.9\textwidth\epsfysize=0.7\textwidth\epsfbox[-10 -10 530 230] 
{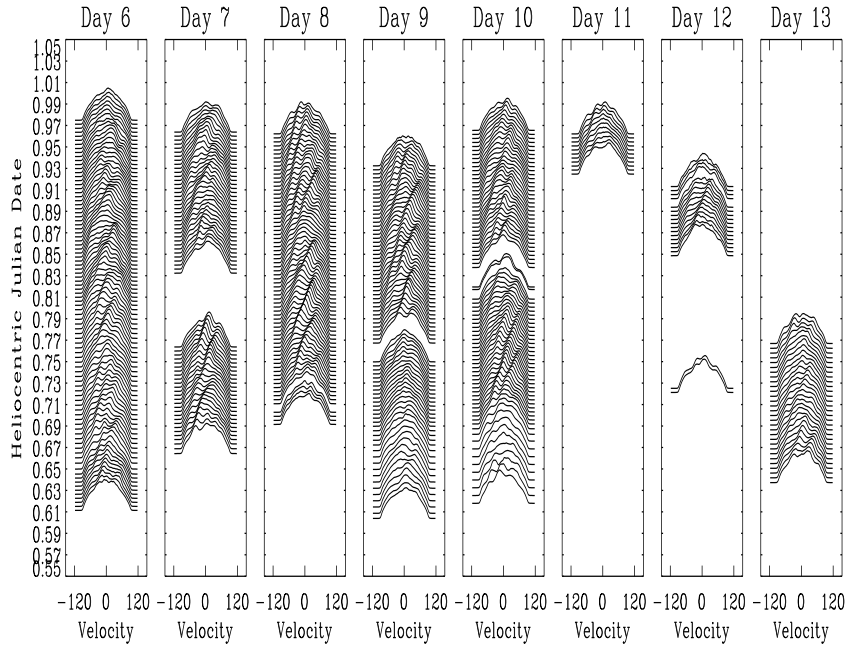}} 
\caption{Time series of broadening functions of $\varepsilon\,$Cep 
using the DDLPV 
technique (see text for an explanation). The profiles are plotted as a function 
of Doppler shift and time. Patterns of variation owing to the pulsations of the 
star can be seen as bumps which travel through the profiles. Figure taken from 
Kennelly et al. (1998a) with permission.} 
\end{figure}

The only assumption
that Kennelly et al.\ (1998a)
 use is that the rotation causes the bump motion (i.e., $v \ll 
v\sin\,i$) and that an accurate estimate of the rotational velocity is known.
We show in Figure\,8 the two-dimensional amplitude spectrum of 
$\varepsilon\,$Cep,
obtained from Fourier Doppler Imaging of the time-variable component of the
broadening function in Doppler space and time. The measured frequencies are
indicated as crosses (Kennelly et al.\ 1998a).
\begin{figure}
\mbox{\epsfxsize=0.9\textwidth\epsfysize=0.7\textwidth\epsfbox[-10 -10 530 530]
{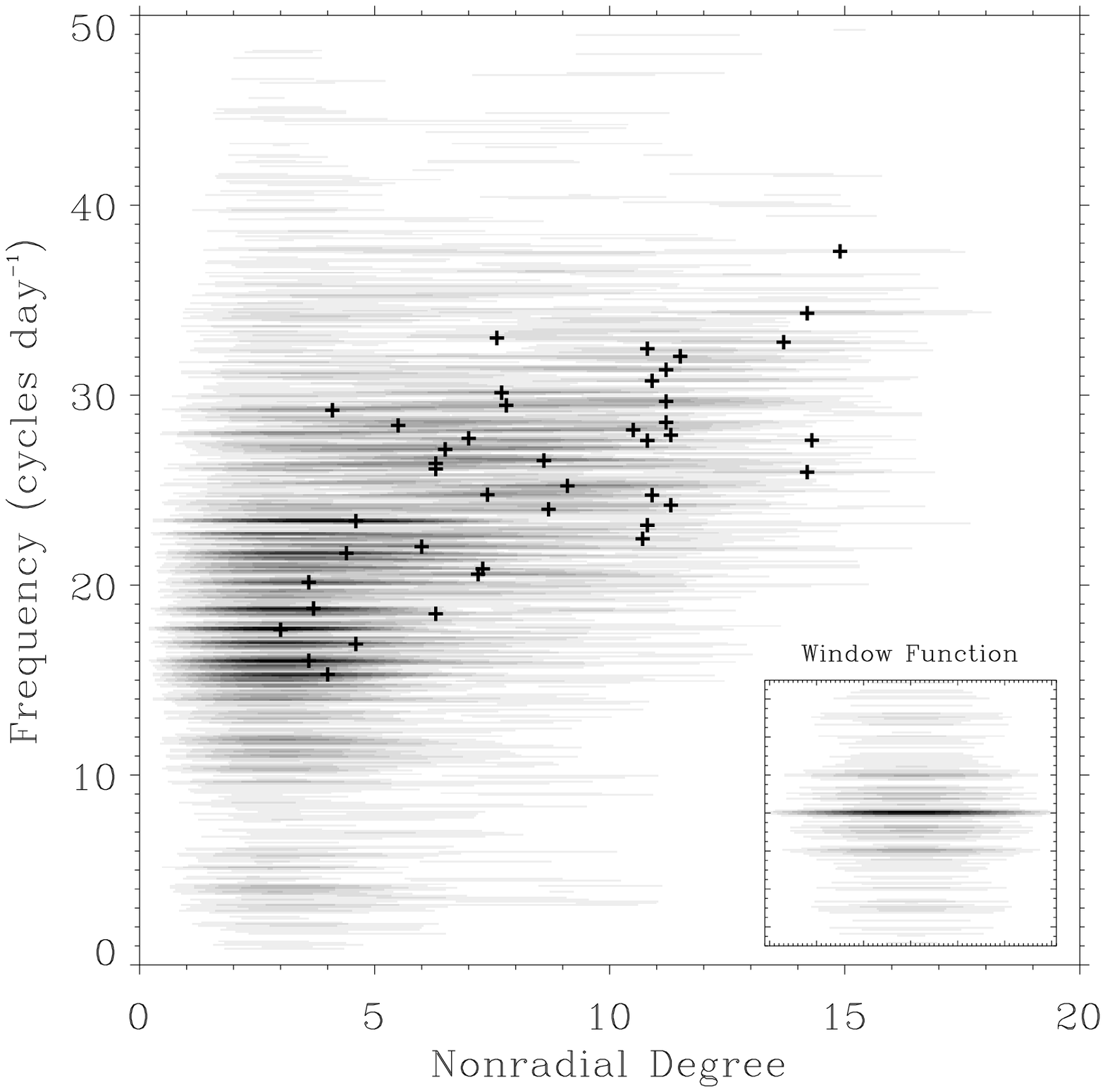}}
\caption{The two-dimensional Fourier amplitude spectrum and frequency analysis
of $\varepsilon\,$Cep. 
The spectrum was computed using the Fourier Doppler Imaging technique (see text
for an explanation). The window function for the spectrum is illustrated as an
inset. The frequencies identified by the two-dimensional frequency analysis are
indicated as crosses. Figure taken from Kennelly et al.\ (1998a) with
permission.}
\end{figure}

The Doppler Imaging technique is not really suited to analyse data sets that
contain very few spectra per night. In this sense, its applicability is limited
to short-period pulsators (i.e., $p$-mode pulsators) for which one usually
focuses on one or a few stars per night during an observing mission with a
time base of typically a week. The observing strategy with long-term
spectroscopy, which is necessary to analyse the line-profile variations of
$g$-mode pulsators, is totally different. In this case one takes a large sample
of stars which are each measured between two and five times per night during
weeks that are in their turn separated by months (for an example of a long-term
spectroscopic project, see Aerts et al.\ 1999a).  Grey-scale representations and
identification methods as the ones shown in Figure\,5 and
Figure\,6 become meaningless in this case, the more so because most
$g$-mode pulsators found up to now are slow rotators.

THE major problem with the Doppler Imaging technique, in whatever form, is that
the spherical wavenumbers are estimated from diagnostics that are not
immediately interpretable in terms of the physics involved in the pulsational
displacement. Indeed, the underlying mathematical basis for this method is
lacking. A first effort to link the physical quantities directly to the
amplitude and phase in Fourier space was undertaken by Hao (1998). This effort
did not lead to new results compared with those already obtained by Telting \&
Schrijvers (1997) from their simulation study and does not give any information
on the velocity parameters other than the degree of the pulsation.  In fact,
only one, and in the best case the two, wavenumber(s) is (are) estimated as
real number(s) from the observed phase changes. A real value of $\ell$ and $m$
has, however, no physical meaning.  Moreover, no information can be derived,
for example, for
the amplitude of the pulsation and for the inclination angle. On the
other hand, multiperiodicity is easily taken into account, contrary to the
other methods. We therefore advise to combine Doppler Imaging with line-profile
fitting once the best estimates of $(\ell,m$) for each of the modes are
obtained. We finally recall that the method is only applicable to rapid
rotators because of the basic assumption that the rotation carries bumps across
the profiles.  For the same reason it is also unsuitable to detect axisymmetric
modes ($m=0)$ and low-degree tesseral modes.

\subsection{The moment method}
As an alternative to the line-profile-fitting technique, Balona
(1986a, b; 1987; 1990) proposed a new method to identify
the modes from line-profile variations:~{\it the moment method}. This method is
based on the time variations of the first few moments of a line profile. We
have extended this method and applied it for the first time to line-profile
variations of a real star, namely the monoperiodic $\beta\,$Cephei star
$\delta\,$Ceti (Aerts et al.\ 1992). In the meantime, this method turned out to
be the best identification technique for slow rotators.  
We here briefly sketch the basic ideas of
the moment method in our formulation (see Aerts 1996 for the latest version).

Since a line profile is a convolution (see Equation\,(\ref{conv})) of an
intrinsic profile (here denoted by $g(v)$) and the intensity in the direction
of the observer integrated over the visible surface (denoted by $f(v)$), the
$n$th moment of a line profile $(f\ast g)(v)$ is defined as
\begin{equation}
\label{defmom}
<v^n>_{f\ast g}\equiv {\displaystyle{\int_{-\infty}^{+\infty}}v^n f(v)\ast
g(v)\,dv \over
\displaystyle{\int_{-\infty}^{+\infty}}f(v)\ast g(v)\,dv},
\end{equation}
where $v$ is the total velocity component in the line of sight.  In principle,
all the moments are needed to give a complete description of the line profile,
but we have shown that the first three moments contain enough information to
accurately describe the profiles (Aerts et al.\ 1992, De Pauw et al.\ 1993).
Note that normalised moments are considered such that they are only slightly
influenced by temperature variations and by uncertainties in the intrinsic
profile.

We have shown that, in the slow-rotation approximation, the first three moments
of a monoperiodic pulsation with frequency $\omega$ are given by~:
\begin{equation}
\label{mono1}
<v>_{_{f\ast g}}=v_{\rm p}A(\ell,m,i)\sin[(\omega-m\Omega)t+\psi],
\end{equation}
\begin{equation}
\label{mono2}
\renewcommand{\arraystretch}{1.5}\begin{array}{ll}
<v^2>_{_{f\ast g}}=\!\!\!&v_{\rm p}^2C(\ell,m,i)
{\sin[2(\omega-m\Omega)t+2\psi+\frac{3\pi}{2}]}\\ &+v_{\rm
p}v_{_{\Omega}}D(\ell,m,i){\sin[{(}\omega-m\Omega)t+\psi+\frac{3\pi}{2}]}\\
&+v_{\rm p}^2E(\ell,m,i)+\sigma^2+b_2v_{_{\Omega}}^2,\end{array}
\end{equation}
\begin{equation}
\label{mono3}
\renewcommand{\arraystretch}{2}\begin{array}{ll}
<v^3>_{_{f\ast g}}=\!\!\!&v_{\rm p}^3F(\ell,m,i)
\sin[3(\omega-m\Omega)t+3\psi]\\
&+v_{\rm
p}^2v_{_{\Omega}}G(\ell,m,i){\sin[2{(}\omega-m\Omega)t+2\psi+\frac{3\pi}{2}]}\\
&+\left[v_{\rm p}^3R(\ell,m,i)+v_{\rm p}v_{_{\Omega}}^2S(\ell,m,i)+v_{\rm
p}\sigma^2T(\ell,m,i{)}
\right]\\&\times \, {\sin}[{(}\omega-m\Omega)t+\psi]
\end{array}\end{equation}
(Aerts et al.\ 1992).  In these expressions, $\psi$ is a phase constant
depending on the reference epoch, $i$ is the inclination angle between the
rotation axis and the line of sight, $v_{_{\Omega}}$ is the projected rotation
velocity (a uniform rotation is assumed), $b_2$ is a constant depending on the
limb-darkening function, $\sigma$ again represents the width of the Gaussian
intrinsic profile as in Section\,3, and the functions $A,C,D,E,F,G,R,S,T$
depend on the kind of mode and on the inclination. They contain the complete
physics of the pulsation.  For an explicit expression of these functions, we
refer to Aerts et al.\ (1992), but we point out here that these functions are
the same for positive and negative azimuthal numbers because the slow-rotation
approximation is used.  It is then impossible in this description to decide
from the moments how a mode travels with respect to the rotation.

\begin{figure}
\mbox{\epsfxsize=0.9\textwidth\epsfysize=0.9\textwidth\epsfbox[50 20 700 560]
{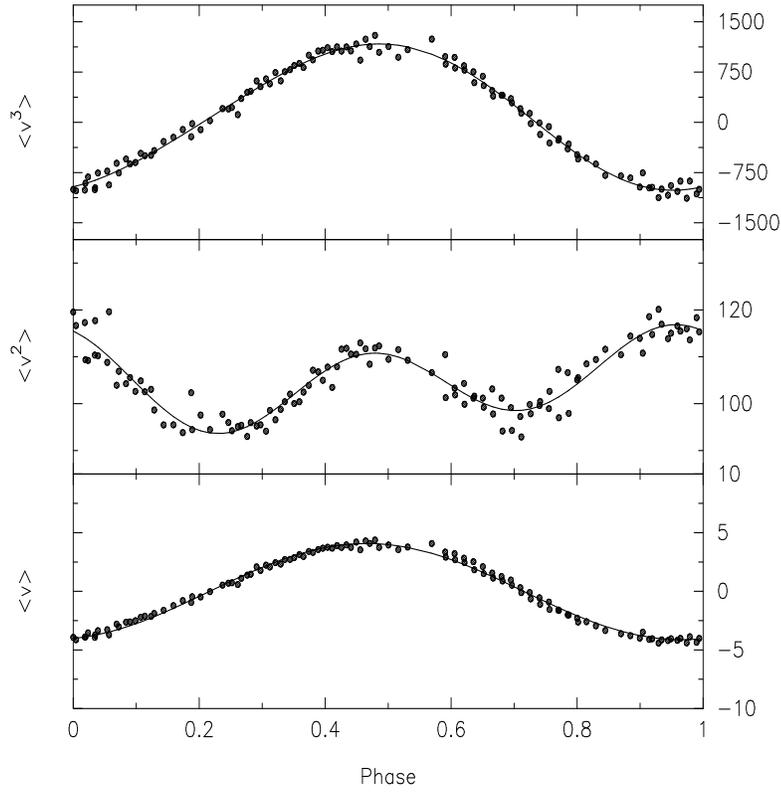}}
\caption{The three observed (dots) velocity moments of the 
Ca\,I\,$\lambda\lambda\,6122.21$\AA\ line of $\rho\,$Puppis. The first, second,
and third moments are expressed in respectively km/s, (km/s)$^2$, and
(km/s)$^3$. The full line is a
fit for a monoperiodic pulsation model for the frequency $f=7.098168\,$c/d. We
refer to Mathias et al.\ (1997) for a complete description of the data.}
\end{figure}

By means of an example, we show in Figure\,9 the first three moments of the
Ca\,I\,$\lambda\lambda\,6122.21$\AA\ line observed for the $\delta\,$Scuti star
$\rho\,$Puppis.  The observed moments are fitted with a monoperiodic pulsation
model for the frequency $f=7.098168\,$c/d (for a full description of the data,
see Mathias et al.\ 1997). It is noted from the middle panel of this figure
that the second moment of $\rho\,$Puppis is dominated by the frequency
$2\omega$, a situation that is typical in the case of an axisymmetric mode (see
Aerts et al.\ 1992).  The top panel of the figure shows that $<v^3>$ is
dominated by the frequency $\omega$. This is a general characteristic of the
third moment since the term varying with $\omega$ is influenced by all
velocities together, i.e., by the rotation, the pulsation, and the intrinsic
profile, while this is not the case for the other two terms (see
Expression\,(\ref{mono3})).

The periodograms of the three moments can immediately be interpreted in terms
of the periods that are present, while the corresponding phase diagrams of the
moments are interpretable in terms of all the non-radial pulsation parameters.
The basic idea is to compare the observed variations of the moments with
theoretically calculated expressions for these variations for various
pulsation modes, and so to determine the mode that best corresponds to the
observations.  This is achieved through the construction of a so-called
discriminant, which is based on the amplitudes of the moments:
\renewcommand{\arraystretch}{2.5}
\begin{equation}
\begin{array}{ll}
\label{discr}
\Gamma_{\ell}^m&(v_{\rm p},i,v_{_{\Omega}},\sigma)\equiv
\Biggl[
\biggl|AA-v_{\rm p}|A(\ell,m,i)|f_{AA}\biggr|^2\\
&+\biggl(\biggl|CC-v_{\rm p}^2|C(\ell,m,i)|\biggr|^{1/2}f_{CC}\biggr)^2\\
&+\biggl(\biggl|DD-v_{\rm p}v_{_{\Omega}}|D(\ell,m,i)|\biggr|^{1/2}f_{DD}
\biggr)^2\\
&+\biggl(\biggl|EE-v_{\rm
p}^2|E(\ell,m,i)|-\sigma^2-b_2v_{_{\Omega}}^2\biggr|^{1/2} f_{EE}\biggr)^2\\
&+\biggl(\biggl|FF-v_{\rm p}^3|F(\ell,m,i)|\biggr|^{1/3}f_{FF}\biggr)^2\\
&+\biggl(\biggl|GG-v_{\rm p}^2v_{_{\Omega}}|G(\ell,m,i)|\biggr|^{1/3}
f_{GG}\biggr)^2\\ &+\biggl(\biggl|RST-v_{\rm p}^3|R(\ell,m,i)| -v_{\rm
p}v_{_{\Omega}}^2|S(\ell,m,i)|\\ &\ \ \ \ \ \ -v_{\rm
p}\sigma^2|T(\ell,m,i)|\biggr|^{1/3}f_{RST}\biggr)^2
\Biggr]^{1/2}.
\end{array}
\end{equation} 
The functions $f_{AA},\ldots,f_{RST}$ are weights given according to the
quality of the fits to the moments. We refer to Aerts (1996) for their
calculation and for a more detailed description and an evaluation 
of the discriminant.

To define a criterion for mode identification, we proceed as follows. The
function $\Gamma_{\ell}^m(v_{\rm p},i,v_{_{\Omega}},\sigma)$ is minimised for
each set of values $(\ell,m)$~:
\begin{equation}
\gamma_{\ell}^m \equiv \ds{\min_{v_{\rm p},i,v_{_{\Omega}},\sigma} 
\Gamma_{\ell}^m(v_{\rm p},i,v_{_{\Omega}},\sigma)}.
\end{equation}
The ``best solution'' for $\ell$ and $m$ is defined as the one for which
$\gamma_{\ell}^m$ attains the lowest value; it then also provides values for
$v_{\rm p}, i, v_{_{\Omega}}$ and $\sigma$.

Our discriminant was thoroughly tested (Aerts 1996) and turned out to be more
accurate compared to the one presented by Balona (1990), which is based on the
first two moments only.  As with line-profile fitting, both the wavenumbers
($\ell,m)$ and all the other velocity parameters are derived.  The moment
method is particularly suited to identify lower-degree modes ($\ell\leq\,6$) in
slow rotators. In this sense, it is completely complementary to the Doppler
Imaging method.  The reason for this limitation is that it uses integrated line
profiles, because of which high-degree modes are almost completely canceled
out in the moment variations.  The
code that calculates the minima of the discriminant as presented here is
written in the statistical package GAUSS and is available upon request from the
first author of this paper.

We recall that the discriminant is unable to find the sign of $m$, because the
theoretical expressions for the moments have only been determined in the case
that the Coriolis force can be neglected. A generalisation that includes the
Coriolis force, and thus is able to derive the sign of $m$,  has been done as
well (Aerts, unpublished).

A generalisation of the moment method to multiperiodic pulsations has been
proposed (Mathias et al.\ 1994a). From our study and the one by Aerts et al.\
(1994b) it is clear that the moment method is less accurate for multiperiodic
stars, but still better than any other alternative in the case of slow
rotation. The biggest problem in the treatment of multiperiodic variations is
the appearance of long beat periods due to the interaction of the different
modes. This effect requires  many observations, well-spread over the total
beat period. A second theoretical problem is that a discriminant constructed to
identify all the present modes at the same time is numerically too involved to
be of any practical use. We are thus obliged to identify each mode separately
by means of the discriminant given in Expression\,(\ref{discr}). In this way,
all the information present in the beat-terms is lost.

An application of the discriminant to the moments of $\rho\,$Puppis shown in
Figure\,9 is given in Table\,2.
\renewcommand{\arraystretch}{1.1}
\begin{table}
\caption{The minima of the discriminant for the main mode ($f=7.098\,$c/d) 
of $\rho\,$Pup}
\begin{center}
\begin{tabular}{ccccccc}
\tableline
$\ell$ & $|m|$ & $\gamma_{\ell}^m$ & $v_{\rm p}$ & $i$ & $v\sin\,i$ &
$\sigma$\\
\tableline
0&0&0.08&5.6&--&15.3&6.5\\
1&1&0.13&10.0&$38^{\circ}$&14.8&5.9\\ 
2&1&0.17&12.1&$64^{\circ}$&16.4&2.2\\
1&0&0.18&5.0&$7^{\circ}$&19.6&1.7\\ 
2&2&0.23&15.0&$53^{\circ}$&10.3&4.8\\
$\vdots$&$\vdots$&$\vdots$&$\vdots$&$\vdots$&$\vdots$&$\vdots$\\
\hline
\end{tabular}
\end{center}
\end{table}
Clearly, the main mode of $\rho\,$Puppis is a radial one with a pulsation
velocity amplitude of some 6\,km/s.  As already mentioned in
Section\,4.1, we found evidence of two additional frequencies in
our data: 7.82\,c/d \& 6.31\,c/d (Mathias et al.\ 1997). Their amplitudes are
too low to achieve an unambiguous mode identification. They are not found in
photometry so far.

Up to now, the moment method in the given formulation has mainly been applied
to $\beta\,$Cephei stars (see e.g., Aerts et al.\ 1992, Mathias et al.\
1994a,b, Aerts et al.\ 1994a,b) but also to two $\delta\,$Scuti stars (20\,CVn:
Mathias \& Aerts 1996, $\rho\,$Puppis: Mathias et al.\ 1997).  Previous
attempts to identify modes in $\delta\,$Scuti stars with Balona's (1990)
version of the moment method are presented by Mantegazza et al.\ (FG\,Vir:
1994) and by Mantegazza \& Poretti (X\,Caeli: 1996). We have recently obtained
a large data set of high-quality line-profile variations of 20\,CVn to check
our findings presented in the 1996 paper, which were based on only very few
spectra. We will proceed with the reduction and analysis process of the new
data sets in the forthcoming months (Mathias et al., in preparation).

THE major limitation of the moment method is the fact that no confidence
intervals for the minima of the discriminant and the corresponding velocity
parameters $v_{\rm p}, i, v_{_{\Omega}}, \sigma$ are available. Therefore, the
competing modes as listed in Table\,2 are difficult to compare with each other.
The standard error of the minimum and of the estimates for $v_{\rm p}, i,
v_{_{\Omega}}$ and $\sigma$ is caused by observational noise, by limitations of
the model describing the line-profile variations due to non-radial pulsation,
and also by numerical inaccuracies occurring in the determination of the
moments, of the amplitudes of the moments, and of the minima of the
discriminant. Unfortunately, no method is found up to now to determine these
uncertainties. We are currently elaborating on a statistically founded method
to try and estimate these standard errors.  If we succeed in doing so, then the
major drawback of this method will be overcome.  Again, line-profile fitting
for the best solutions found by the discriminant is helpful to check the result
of the mode identification.
\vspace{-1.0em}

\section{Comparison between the methods}

\renewcommand{\arraystretch}{1.1}
\begin{table}[b!]
\vspace{-1em}
\caption{The main properties of each of the three identification methods
based on observed line-profile variations. LPF, DI, and MM stand for
respectively line-profile fitting, Doppler Imaging, and the moment method.}
\begin{center}
\tabcolsep=10pt
\begin{tabular}{rccc}
\tableline
& LPF & DI & MM\\\tableline
Deduced parameters & $\ell, m$, ampl & $\ell, m ? $ & $\ell, m$, ampl\\
& vsini, $\sigma, i$ & &vsini, $\sigma, i$\\\tableline
Limitations & no & v/vsini$\leq\,20\%$ & $\ell\leq\,6$ 
\\\tableline
Multiperiodicity & no & easy & possible \\\tableline
Underlying physics & yes & no & yes \\\tableline
Standard errors & no & no & not yet \\\tableline
Computation time & long & short & in between \\\tableline
Additional modeling & no & necessary & as check 
\\\tableline
\end{tabular}
\end{center}
\end{table}

We have already mentioned the main advantages and disadvantages for each of the
three methods described above. We briefly review them in Table\,3.  The
methods are complementary in the sense that one is suited for slow rotators
with low-degree modes (moment method), another for rapid rotators with
high-degree modes (Doppler Imaging) and the third (line-profile variation
fitting) is very useful (moment method)/necessary (Doppler Imaging) 
as a check of the results obtained with the other two methods.

\section{Conclusions and future developments}

In this review, we have discussed the different mode-identification techniques
that are currently used to study the non-radial pulsations in pulsating stars
from observations of line-profile variations. Three basic methods are
presented, are compared to each other and applications to real observations of
$\delta\,$Scuti stars are described.

Line-profile variations offer a very detailed picture of the various aspects of
the pulsation velocity field. On the other hand, photometric observations are
easier to obtain on a long time-basis and are as such often superior for
a very accurate determination of the pulsation frequencies, especially in the
case of lower-degree modes. High-degree modes hardly show up in photometry and
can only be found from high-quality line-profile variation data. An example of 
additional modes being seen in spectra compared to photometry is presented by De
Mey et al.\ (1998). They have analysed high-quality line profiles of the
multiperiodic 
$\delta\,$Scuti primary of the double-lined spectroscopic binary $\theta\,$Tuc.
The dominant frequency is the same in the photometric and spectroscopic data,
but the second frequency that shows up in the spectra was never found before in
photometry. This example confirms that 
the gathering of simultaneous photometry and spectroscopy is the best strategy
to find a complete and accurate identification of all the appearing modes
in multiperiodic stars.

Future possible improvements from a theoretical point of view concern on the
one hand the development of mathematical expressions for the phase and
amplitude in Fourier space in such a way that these quantities can be
immediately interpreted in terms of the physical parameters of the pulsational
velocity field. We also briefly mention that a new method of ``Doppler
Mapping'' was recently presented by Berdyugina et al.\ (2000). They apply a
spectral inversion technique to obtain maps of the surface corotating with the
dominant pulsation mode. From these maps, they determine the pulsation degree
and study the latitudinal distribution of the pulsation field.  The method
still needs to be further explored. Secondly, the inclusion of temperature
variations during the pulsation cycle is still not accurately done, since an
adiabatic pulsation is assumed while it is to be expected that non-adiabatic
effects are important in the outer region of the atmosphere where the observed
spectral lines are formed. Finally, the inclusion of centrifugal forces may be
an improvement for the most rapid rotators. The latter is only necessary for
stars rotating close to their break-up velocity.

From an observational point of view, large progress can be expected in the near
future now that better and better detectors become available. For the
short-period $\delta\,$Scuti stars, the major problem in obtaining high
temporal and spatial resolution profiles is that the ratio of the integration
time to the main pulsation period is rather high. This was one of the reasons
why the application of the moment method to the stars FG\,Vir and X\,Caeli was
not very successful. The abovementioned ratio in these cases was respectively
13\% and 8\%, while it amounts to only 1\% for our profiles of $\rho\,$Puppis
shown in Figure\,4.  For such high ratios, the pulsational motion is averaged
out over a part of the cycle and this prevents unambiguous identifications,
especially for multiperiodic stars.

An interesting new technique for the interpretation of  
line-profile variations is by working
with cross-correlation functions instead of real spectra. Such a technique can
be performed by means of current spectrographs such as ELODIE attached to the
1.93m telescope in the Haute-Provence Observatory. Our analysis of 20\,CVn
(Mathias \& Aerts 1996) was already based on cross-correlation profiles and has
shown that they perfectly contain the pulsational motion on the condition that
the correlation is based on a suitable set of selected spectral lines.  
By using a cross-correlation function, one can significantly decrease the
integration time and still obtain a high S/N ratio. At the same time, one can
observe optically fainter stars with success.  More accurate versions of
ELODIE-type spectrographs are CORALIE, attached to the Swiss 1.2m telescope and
FEROS, attached to the ESO 1.5m telescope, both situated at La Silla in Chile.

Finally, we would like to point out that mode identification from line-profile
variations will become an important tool to obtain some information on the
nature of the excited modes in stars belonging to the new class of
$\gamma\,$Dor stars. For reviews on this new group of pulsating stars we refer
to Krisciunas (1998) and to Zerbi (these proceedings).  Since the multiperiodic
variations detected in them have periods roughly a factor 20 longer than the
period of the radial fundamental mode for such stars, high-order $g$-modes are
believed to be the cause of the variability. However, there is yet no pulsation
mechanism that can explain the onset and the maintenance of the pulsations in
these stars.

Handler \& Krisciunas have given subsequent updated lists of {\it bona fide}
members of the group, which currently constitutes 13 members.  We have recently
taken the first steps towards the discovery of cool 
$g$-mode pulsators by searching
for $\gamma\,$Dor stars in an unbiased sample of 39 new variable A2--F8 stars
discovered by means of the Hipparcos mission (Aerts et al.\ 1999b).  We have
reported the discovery of 14 new $\gamma\,$Doradus variables among this
unbiased sample.  We primarily focussed on the limited group of new variables
for which both Hipparcos and Geneva data are available, mainly because the
latter allow an accurate determination of the effective temperature. It is very
likely, however, that our more extended list of 200 unclassified variable
A2--F8 stars of which no Geneva data are at our disposal contains more objects
of this type.  This seems to be confirmed by a recent analysis by Handler
(1999).

In 1996, we also started a search for new $\gamma\,$Dor stars by means of
ground-based Geneva photometry. Our search has resulted so far in the discovery
of three new and some five suspected $\gamma\,$Dor stars (Eyer \& Aerts, 
2000).  In order to firmly establish the $\gamma\,$Dor nature of all
these new candidates we have started a long-term spectroscopic campaign with
CORALIE in the course of 1997, which is still ongoing. We found line-profile
variability from our high-resolution spectra for almost all candidates. Some of
them, however, turn out to be binaries (Eyer \& Aerts, in preparation).

Line-profile studies of $\gamma\,$Dor stars are still scarce. Examples in which
a large amount of spectra have been obtained and analysed are given by Balona
et al.\ (1996, $\gamma\,$Dor) and by Aerts \& Krisciunas (1996, 9\,Aur). The
latter study is based on cross-correlations obtained with the (by now
unmounted) spectrograph CORAVEL and showed convincingly that such correlation
functions indeed contain a sufficient amount of information to characterise the
pulsational behaviour.

It is clear that a combination of long-term photometry and spectroscopy is
essential and the only way to study the multiperiodic variability in the
$\gamma\,$Dor stars. The best observing strategy is the same as the one for the
slowly pulsating B stars (Aerts et al.\ 1999a), which are also multiperiodic
$g$-mode pulsators. At present, we do not yet have a sufficient amount of line
profiles for our targets, but we will continue our monitoring of $\gamma\,$Dor
stars with CORALIE during the forthcoming years. This will eventually lead to
mode identifications, by applying the moment method.

\end{document}